\documentclass[12pt]{article}

\usepackage{amsfonts,amsmath}
\usepackage{latexsym}

\usepackage{epsfig}
\def\void{}
\def\labelmark{}

\newenvironment{formula}[1]{\def\labelname{#1}
\ifx\void\labelname\def\junk{\begin{displaymath}}
\else\def\junk{\begin{equation}\label{\labelname}}\fi\junk}%
{\ifx\void\labelname\def\junk{\end{displaymath}}
\else\def\junk{\end{equation}}\fi\junk\labelmark\def\labelname{}}

{\ifx\void\labelname\def\junk{\end{array}\end{displaymath}}
\else\def\junk{\end{array}\right.\end{equation}}
\fi\junk\labelmark\def\labelname{}\def\junk{}
\def\arraystretch{1}}

\newcommand{\beq}{\begin{formula}}
\newcommand{\eeq}{\end{formula}}
\newcommand{\beqv}{\begin{formula}{}}

\newcommand{\rf}[1]{(\ref{#1})}
\newcommand{\oh}{\frac{1}{2}}

\newcommand{\bea}{\begin{eqnarray}}
\newcommand{\eea}{\end{eqnarray}}
\newcommand{\beas}{\begin{eqnarray*}}
\newcommand{\eeas}{\end{eqnarray*}}
\newcommand{\beqs}{\begin{displaymath}}
\newcommand{\eeqs}{\end{displaymath}}

\newcommand{\br}{\langle}
\newcommand{\kt}{\rangle}

\newcommand{\ep}{\varepsilon}

\newcommand{\cC}{{\cal C}}

\newcommand{\cA}{{\cal A}}

\newcommand{\cB}{{\cal B}}

\newcommand{\ben}{\begin{equation}}
\newcommand{\een}{\end{equation}}

\newcommand{\bdm}{\begin{displaymath}}
\newcommand{\edm}{\end{displaymath}}

\newcommand{\pa}{\partial}

\begin{document}
 \topmargin 0pt
 \oddsidemargin 5mm
 \headheight 0pt
 \topskip 0mm

 \addtolength{\baselineskip}{0.4\baselineskip}

 \pagestyle{empty}

 \vspace{0.1cm}

\hfill

\vspace{1cm}

\begin{center}

\medskip

{\Large \bf Random walks on combs}

\medskip

\vspace{1.2 truecm}

 \vspace{0.7 truecm}
{\bf Bergfinnur Durhuus}\footnote{durhuus@math.ku.dk}

\vspace{0.5 truecm}

Matematisk Institut, Universitetsparken 5

2100 Copenhagen \O, Denmark

 \vspace{.8 truecm}

{\bf Thordur Jonsson}\footnote{
thjons@raunvis.hi.is}
\footnote{Permanent address: 
University of
Iceland, Dunhaga 3,
107 Reykjavik, Iceland} ~and ~
{\bf John F. Wheater}\footnote{j.wheater1@physics.ox.ac.uk}

\vspace{.5 truecm}  

Theoretical Physics, University of Oxford 

1 Keble Road, OX13NP, UK

 \vspace{1.5 truecm}

 \end{center}

 \noindent
 {\bf Abstract.} We develop techniques to obtain rigorous bounds on the
behaviour of random walks on combs.   Using these bounds we calculate exactly 
the spectral dimension of
random combs with infinite teeth at random positions or teeth with random
but finite length.  We also calculate exactly
 the spectral dimension of some fixed
non-translationally invariant combs.  We relate the spectral dimension to the
critical exponent of the mass of the two-point function for random walks on
random combs, and compute mean displacements as a function of walk duration.
We prove that the mean first passage time is generally infinite for combs with
anomalous spectral dimension.

 \vfill

 \newpage
 \pagestyle{plain}

\section{Introduction}
The fractal structures of random geometrical objects have been under
intensive investigation 
for a number of years, both in connection with quantum
gravity \cite{book} and in the study of disordered materials 
\cite{bookben,haus,revfractal}.   This work is to a large extent aimed at
understanding the geometric characteristics of generic objects in the
ensembles under study and how these characteristics 
are reflected in physical phenomena.

An important notion in the study of fractal geometries is
the concept of dimension.  Definitions of dimension 
which agree on smooth manifolds do in general
not do so in the random or fractal case.
One important concept is that of {\it spectral dimension} 
which is defined to be
$d_s$ provided the heat kernel at coinciding points, averaged over the
random geometries and viewed as a function of time $t$,  decreases as
$t^{-d_s/2}$ as $t\to\infty$.   Equivalently, the spectral dimension is a
measure of how likely a random walker is to be at the starting point after
time $t$.  This notion of dimension is in general
different from that of Hausdorff dimension $d_H$ 
which is defined in terms of the
growth of the expectation value of the 
volume of a geodesic ball of radius $r$ as $r\to\infty$:
\beq{1}
\br  B(r)\kt\sim r^{d_H}.
\eeq
We will study in detail many examples of this 
for the case of random combs in this
paper.  The discrepancy between these dimensions is also well demonstrated, at
least numerically, in quantum gravity.

Early work on the spectral dimension in quantum gravity was done by
numerical simulation \cite{earlynumsim}. This lead to the investigation of random walks on random trees and the spectral dimension of random
trees was calculated analytically in \cite{tjjw};  the extension to non-generic trees was given in \cite{jwc}. In \cite{scaling}
a scaling relation was derived which relates the spectral dimension to the
extrinsic Hausdorff dimension.    Related work on the spectral dimension of
trees can
be found in \cite{italianpaper,donetti2}.  
In the condensed matter community the
spectral dimension has been investigated for a variety of systems, see, e.g., 
\cite{bookben}.  While
very few exact results have been obtained, random combs in particular have
been studied numerically as well as by mean field theory methods.  Mean
field theory simplifies the problem since it allows one to model the
walk on a comb by a random  walk on the spine of a comb with a waiting
time distribution which is taken to be the same for all vertices on the spine.
The waiting times arise from the excursions that the walk makes into the
teeth of the comb.
The spectral dimension of random combs with teeth whose
lengths obey a power law distribution has been studied in 
\cite{powerlaw}, see also \cite{physica}.  If the exponent
of the power law distribution is $a$ then the spectral dimension was 
found to be given by
\beq{2}
d_s= {4-a\over 2}
\eeq
for $a\leq 2$ but $1$ otherwise.    In this paper 
we prove this result which
shows that mean field theory is exact in this case.  
Mean field theory was shown to be exact in a special case in \cite{mft}.
We will also show that the spectral dimension of random combs 
whose teeth may be
infinitely long is always $3/2$.

We develop technical tools to prove the above mentioned results and also
apply those techniques to random trees and to 
some examples of non-random combs.   The main new idea is a splitting of
random walks as well as random combs into subsets that either yield
exponentially suppressed or uniformly controllable contributions to the
quantities under consideration (typically $d_s$).  The tools are
recursion relations for generating functions, simple monotonicity
results and convexity arguments.  We believe that 
the methods can be applied to study random walks on more
complicated random graphs.

In the next section we define the random comb ensembles we wish to study and
the most important generating functions and critical exponents.  We
establish simple monotonicity results and use them to obtain some
elementary bounds.
In Section 3 we study random combs which have an infinitely long tooth at
each site on the spine with a nonzero probability.  In this case the
spectral dimension is always $3/2$ which is the same as the spectral
dimension for a comb with all teeth infinitely long.  In Section 4 we
calculate the spectral dimension of combs with random but 
finitely long teeth and show that the spectral dimension is determined by the
tail of the length distribution.  In Section 5 we apply our methods to prove
upper and lower bounds on the spectral dimension of random trees.  In
Section 6 we calculate the spectral dimension of fixed combs whose
toothlength 
increases along the spine and also combs with infinite teeth whose
separation increases along the spine.
Section 7 contains results about transport along the backbone of
the combs and the full heat kernel on random combs.  In the final section we
discuss the relevance of our methods and results for random geometry in
general and
compare our work with relevant results in 
the mathematics literature.  
Various technical calculations are relegated to two appendices.

\section{Preliminaries}
Let $N_\infty$ denote the nonnegative integers regarded as a graph so that $n$
has the neighbours $n\pm 1$ except for $0$ which only has $1$ as a neighbour.
Let $N_\ell$ be the integers $0,1,\ldots ,\ell$ regarded as a graph so that
each integer $n\in N_\ell$ has two neighbours $n\pm 1$ except for $0$ and $\ell$
which only have one neighbour, $1$ and $\ell-1$, respectively.  A comb 
$C$ is an
\begin{figure}[thb]
  \begin{center}
      	        \includegraphics[width=10cm]{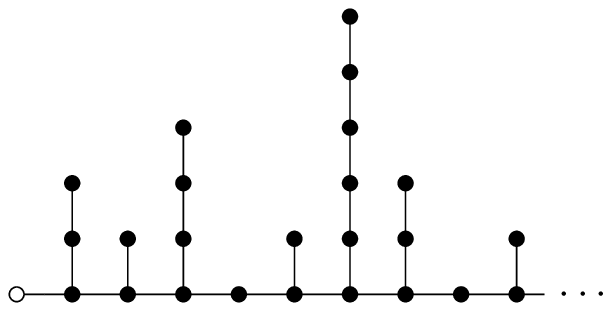}
		    \caption{A comb.}
    \label{fig1}
      \end{center}
      \end{figure}
infinite rooted tree-graph with a special subgraph $S$ called the spine
which is isomorphic to $N_\infty$ with the root at $0$.  At each vertex of
$S$, except the root $0$,  there may be attached one of the graphs $N_\ell$ or
$N_\infty$.  We adopt the convention that these linear graphs which are
glued to the spine are attached at their endpoint $0$.  The linear graphs
attached to the spine are called the teeth of the comb, see Fig.\ 1.
We will find it convenient to say that a vertex on the spine with no tooth
has a tooth of length $0$.  We will denote by $T_n$ the tooth attached to
the vertex $n$ on $S$, and by $C_k$ the comb obtained by removing the links 
$(0,1),\ldots ,(k-1,k)$, the teeth $T_1,\ldots ,T_k$ and relabelling the 
remaining vertices on the spine in the obvious way.

\subsection{Random walks on combs}
We consider simple random walks on the combs.  We assume that the walker
starts at the root unless we specify otherwise.  At each time step the walker
steps with equal probabilities to one of the neighbouring vertices.  This
means that the walker has $1$, $2$ or at most $3$ choices of vertices to
step to at any given time and the corresponding probabilities are
$1$, $1/2$ and $1/3$.  We are interested in the asymptotic properties
of the walk after many time steps.  We regard time as integer valued.  

Given a comb $C$, 
let $p_C(t)$ be the probability that the walker is at the        
root at time $t$. Let $p_C^1(t)$  be the probability that the walker is 
at the root for the first time after $t=0$ at time $t$ with the convention 
$p_C^1(0)=0$.   Clearly both $p_C(t)$ and $p_C^1(t)$ vanish unless $t$ is an
even number.  Given a random walk $\omega$ which comes back to the root 
at time $t$ it is clear that this may be the first return, the second one,
etc.  We can therefore decompose $p_C(t)$ into a sum over walks that have
had a fixed number of intermediate visits to the root before ending there
at time $t$, i.e.,  
\beq{111}
p_C(t)=\delta_{t,0}+\sum_{n=1}^\infty \; \sum_{t_1+t_2+\ldots +t_n=t}\;
\prod_{j=1}^n p_C^1(t_j).
\eeq
We define the generating functions for return to the root and first return
to the root by
\beq{112}
Q_C(z)=\sum_{t=0}^\infty z^tp_C(t)
\eeq
and
\beq{113}
P_C(z)=\sum_{t=0}^\infty z^tp_C^1(t).
\eeq
It follows then from \rf{111} that
\beq{114}
Q_C(z)={1\over 1-P_C(z)}.
\eeq
The function $P_C(z)$ is clearly analytic in the unit disc and satisfies
$|P_C(z)| <1$ for $|z|<1$ .  
Note that functions analogous to $P_C$ and $Q_C$ are defined for
the simple random walk on any rooted graph $\Gamma$ with root of order one.  We
will denote these by $P_\Gamma$ and $Q_\Gamma$.
If $\Gamma$ consists of a single vertex, i.e., $\Gamma=N_0$, 
we adopt the convention
that $P_\Gamma (z)=1$.
We will see in a moment 
that $P_C(1)=1$ for all combs, i.e., the 
simple random walk is recurrent on combs.

Consider a fixed comb $C$.  Any walk that
contributes to $P_C$ can be decomposed 
into a first step from 0 to 1, then an arbitrarily
large number of round trips into the tooth $T_1$ intermingled with round trips 
into the comb $C_1$ and then a final step from 1 to 0.  Each time the walk
is located at 1 the probability of stepping into $T_1$ or $C_1$ is $1/3$ and
likewise the probability of the final step to 0 is $1/3$.  It follows that 
\beq{115}
P_C(z)={z^2\over 3-P_{T_1}(z)-P_{C_1}(z)}.
\eeq
In particular,
if $C$ has no tooth at 1, we have
\beq{116}
P_C(z)={z^2\over 2-P_{C_1}(z)}.   
\eeq
In fact the recurrence relations \rf{115} and \rf{116} are valid for
generalized combs where the teeth are allowed to be arbitrary rooted graphs
(with a root of order 1),
not necessarily the linear ones that we study here.  

Consider now the toothless comb $N_\infty$.  The generating function
for first return to the root, denoted $P_\infty$, satisfies
\beq{117}
P_\infty(z)={z^2\over 2-P_{\infty}(z)}.
\eeq
It is convenient to introduce the variable $x$ related to $z$ by
\beq{118}
1-x=z^2.
\eeq
The generating functions are even functions of $z$ so they can be
regarded as functions of $x$;  we will
denote them by the same symbol which should not cause confusion and
assume from now on that $0\leq x\leq 1$.
From 
\rf{117} we see that
\beq{119}
P_\infty(x)=1-\sqrt{x}.
\eeq
For a finite tooth $N_\ell$ we denote the generating function for first
return of random walks to 0 by $P_\ell$.  An elementary calculation
using the recurrence relation
\rf{116}  and $P_1(x)=1-x$  yields
\beq{120}
P_\ell (x)=1-\sqrt{x}\; {(1+\sqrt{x})^\ell -(1-\sqrt{x})^\ell\over
(1+\sqrt{x})^\ell +(1-\sqrt{x})^\ell},
\eeq
see Appendix 1.
We observe that $P_\ell (x)$ is a decreasing function of $\ell$ for a fixed
$x$ and $P_\ell(x)\to P_\infty (x)$ as $\ell\to\infty$.  

The comb which has an infinite tooth at each vertex on the spine will be
called the {\it full comb} and denoted $*$.  
By  \rf{115} the function $P_*$ satisfies
\beq{121}
P_*(x)={1-x\over 3-P_\infty (x)-P_*(x)}.
\eeq
Using  \rf{119} and the fact that $P_*(x)\leq 1$ we find that
\beq{122}
P_*(x)=1-x^{1/4}\sqrt{1+{5\over 4}\sqrt{x}}+\oh\sqrt{x}.
\eeq
We define the critical exponent $\alpha$ for a comb $C$ by
\beq{123}
1-P_C(x)\sim x^\alpha, ~~~{\rm as}~~x\to 0,
\eeq
where  $f(x)\sim g(x)$ means that for any 
$\ep >0$ there are positive constants $c_1$ and $c_2$ such that              
\beq{123x}      
c_1x^{\ep}f(x)\leq g(x)\leq c_2 x^{-\ep}f(x)      
\eeq
for $0 < x\leq 1$.
We see from \ \rf{119} and \rf{122} that the
half line and the full comb have $\alpha=1/2$ and $\alpha =1/4$, respectively.
It is easy to relate $\alpha$ to the spectral dimension $d_s$.  If 
\beq{124}
p_C(t)\sim t^{-d_s/2}
\eeq
as $t\to\infty$
then
\beq{125}
Q_C(x)\sim x^{-1+d_s/2}
\eeq
as $x\to 0$ and
\beq{126}
 d_s =2-2\alpha , 
\eeq
so the half line and the full comb have spectral dimensions 1 and $3/2$, 
respectively.  The value $d_s=3/2$ for the full comb was first obtained in 
\cite{fullcomb}.  In fact the spectral dimension (if it exists) 
of any comb lies in
the closed interval $[1,3/2]$.  This is a consequence of 
\beq{127}
P_* (x)\leq P_C(x)\leq P_\infty (x)
\eeq
valid for any comb $C$.  The inequalities \rf{127}  follow
from the Monotonicity Lemma below.  Furthermore, the lower bound in 
\rf{127} and \rf{122} imply that random walks on 
combs are recurrent as claimed above.

We note from \rf{115} that, for fixed $x$, $P_C(x)$ is a monotonic 
increasing function of $P_{T_1}(x)$ and $P_{C_1}(x)$.  By applying (7) in
 turn to $P_{C_1}(x), P_{C_2}(x),\ldots P_{C_{k-1}}(x)$  we find by induction 
 the following result:
 
 \medskip
 \noindent
{\bf Lemma A} \emph{The function $P_C(x)$ is a monotonic 
increasing function of
$P_{T_1}(x), \ldots \\ P_{T_k}(x),P_{C_{k}}(x)$ for any $k\ge 1$.}

\medskip

\noindent{\bf Monotonicity Lemma} \emph{$P_C(x)$ is a decreasing 
function of the length,
$\ell_k$, of the tooth $T_k$ for any $k\ge 1$.}

\medskip
\noindent
{\bf Proof} By Lemma A  $P_C(x)$ is a monotonic 
increasing function  of $P_{T_k}(x)$ which is a decreasing
 function of $\ell_k$ according to \rf{120}.

\medskip
\noindent{\bf Rearrangement Lemma} \emph{Let $C'$ be the comb obtained from $C$ by 
swapping the teeth $T_n$ and $T_{n+1}$. Then $P_C(x)>P_{C'}(x)$ if and only if
$ P_{T_n}(x)> P_{T_{n+1}}(x)$.}

\medskip
\noindent
{\bf Proof} By Lemma A, and noting that  $T_1,\ldots T_{n-1}$ 
are the same for $C$ and $C'$, it suffices to prove that 
$P_{C_{n-1}}>P_{C'_{n-1}}$ if and only if $P_{T_n}>P_{T_{n+1}}$. 
By \rf{115} the former holds if and only if 
$ P_{T_n}+ P_{C_n}>P_{T_{n+1}}+ P_{C'_n}$. It is therefore enough to compute
\begin{eqnarray}
\Delta&=&P_{T_n}+ P_{C_n}-P_{T_{n+1}}- P_{C'_n}\cr
&=&(P_{T_n}-P_{T_{n+1}})\left(1-\frac{1-x}{(3-P_{T_{n+1}}-P_{C_{n+1}})
(3-P_{T_{n}}-P_{C_{n+1}})}\right)
\end{eqnarray}
where we have used \rf{115} and the fact that $C_{n+1}=C'_{n+1}$. 
We see that $\Delta>0$ if and only if $P_{T_n}>P_{T_{n+1}}$ which 
completes the proof.

\subsection{The two-point function}

Let $C$ be a comb and let $p_C^1(t;n)$ denote the probability that a random
walk that starts at the root $0$ at time $0$ is at the vertex $n$ on the
spine at time $t$ and has not visited the root in the time 
interval from $0$ to $t$.  We
will refer to the generating function for these probabilities as the
two-point function and denote it by $G_C(x;n)$.  Note that $G_C(x;0)=P_C(x)$
and
\beq{130}
G_C(x;n)=\sum_{t=1}^\infty (1-x)^{t/2}p_C^1(t;n).
\eeq
The two-point function can also be expressed as the sum over all random
walks from $0$ to $n$ which avoid $0$ and end at $n$:
\beq{131}
G_C(x;n)=\sum_{\omega:0\to n,\;\omega_t\neq 0 \;{\rm if} \;t\neq 0}
\prod_{t=0}^{|\omega |-1} \left(\sigma (\omega_t)^{-1}\sqrt{1-x}\right)
\eeq
where $|\omega|$ is the number of steps in the walk $\omega$ and $\sigma
(\omega_t)$ is the order of the vertex $\omega_t$ where $\omega$ is located
at time $t$.  

The representation \rf{131} of the two-point function is quite
useful and allows us to relate it to the first return generating functions
as we now show.
Let $C$ be a comb and let $C_k$ be defined as before.
If we consider a
random walk $\omega$ on $C$ which contributes to the two-point function
$G_C(x;n)$ we can decompose it into a sequence of $n$ random
walks $\omega^1,\ldots , \omega^{n}$ where $\omega^1$ is a  walk from $0$ to
$1$ which is identical to
$\omega$ until $\omega$ leaves the vertex $1$ for the last time before going
to $n$, $\omega^2$ is a walk from $1$ to $2$ which is identical to $\omega$
after it left $1$ for the last time until it leaves $2$ for the last time,
etc. The last walk $\omega^{n}$ is the part of $\omega$ after it left
$n-1$ for the last time and until it ends at $n$.
If for each $k=1,2,\ldots , n-1$
we add a last step to $\omega^k$ back to the vertex $k-1$ and call the
resulting walk $\tilde{\omega}^k$ we see that $\tilde{\omega}^k$ is a walk
from $k-1$ to $k-1$ which contributes to $P_{C_{k-1}}$ and any walk
contributing to $P_{C_{k-1}}$ can arise in this way.  It follows that
for $n>0$
\beq{132}
G_C(x;n)=\sigma (n) (1-x)^{-n/2}\prod_{k=0}^{n-1}P_{C_k}(x),
\eeq
where $\sigma (n)$ is the degree of the vertex $n$.
                                                 
We see from from  \rf{119} and \rf{132} 
that the two-point function for the half line, 
$G_\infty (x;n)$, is given by
\beq{133}
G_\infty (x;n)=2\left({1-\sqrt{x}\over 1+\sqrt{x}}\right)^{n/2}
\eeq
for $n>0$.
We define the mass, $m(x)$, of the two-point function $G_C$ 
by its rate of
exponential decay, i.e.,
\beq{134}
m(x)=-\lim_{n\to\infty}{\log G_C(x;n)\over n}.
\eeq
For an arbitrary comb there is no reason to expect the limit \rf{134} to exist
but the mass associated with the two-point function for the 
half line is clearly
\beq{135}
m_\infty (x)={1\over 2}\log{1+\sqrt{x}\over 1-\sqrt{x}}.
\eeq
We can similarly use  \rf{122} to
compute the two-point function, $G_*(x;n)$ and the mass, $m_*(x)$, for the
full comb.  It furthermore follows from \rf{127} and \rf{132} that  
\beq{137}      
\left({\sigma (n)\over 3}\right)G_*(x;n)\leq 
G_C(x;n)\leq \left({\sigma (n)\over 2}\right)
G_\infty (x;n)      
\eeq
for any comb $C$.
If the mass $m(x)$ exists we define its critical exponent $\nu$ by
\beq{136}
m(x)\sim x^\nu
\eeq
as $x\to 0$.  It is easy to see that $\nu=1/2$ for $m_\infty$ and 
$\nu=1/4$ for $m_*$.
From \rf{137} we conclude that the critical exponent of the mass for any 
comb lies in the
interval $[{1\over 4},\oh ]$.   

The above considerations show that the exponents $\alpha$ and $\nu$ coincide
for the half line and the full comb.
Indeed, we will prove in Section 6 that the scaling relation
\beq{136x}
\alpha =\nu
\eeq
holds quite generally.

\subsection{Random combs}

Let $\cC$ denote the collection of all combs.  Let ${\mathbb Z}_0^+$ denote the
nonnegative integers.  If we are given a probability
measure $\mu$ on ${\mathbb Z}_0^+\cup \{\infty\}$ we can define a probability measure
$\pi$ on $\cC$ by letting the length of the teeth be identically and
independently distributed by $\mu$.  This means that the measure of the set
of combs $\Omega$ with teeth at $n_1, n_2,\ldots ,n_k$ having lengths
$\ell_1,\ell_2,\ldots ,\ell_k$ is
\beq{138}
\pi (\Omega)=\prod_{j=1}^k \mu (\ell_j).
\eeq
We will refer to the set $\cC$ equipped with the probability measure $\pi$
as a random comb.  
Measurable subsets $\cA$ of $\cC$ are called
{\it events} and $\pi (\cA )$ is the probability of the event $\cA$.
We define the first return generating functions for random
walks on random combs as
\beq{139}
\bar{P}(x)=\br P_C(x)\kt
\eeq
where $\br\cdot\kt$ denotes expectation with respect to the measure $\pi$,
i.e.,
\beq{139x}
\br F(C)\kt = \int F(C)\,d\pi
\eeq
for any $\pi$-integrable function $F$ defined on $\cC$.
Similarly,
\beq{140}
\bar{Q}(x)=\br Q_C(x)\kt.
\eeq
If $\bar{Q}(x)\sim x^{-1+d_s/2}$ as $x\to 0$ we say that the spectral
dimension of the random comb is $d_s$.
Similarly we define the exponent $\alpha$ for the random comb by
$1-\bar{P}(x)\sim x^{\alpha}$.  We will see for the examples of random
combs studied in this paper that the relation \rf{126} holds.

The two-point function of the random comb is defined as
\beq{141}
\bar{G}(x;n)=\br G_C(x;n)\kt .
\eeq
We show below that the mass exists for any random comb.
It follows from \rf{137} that 
this mass $\bar{m}(x)$ satisfies the inequalities
\beq{142}
m_\infty (x)\leq \bar{m}(x)\leq m_*(x)
\eeq
and the critical exponent $\nu$ of $\bar{m}$ lies in the interval 
$[{1\over 4}, \oh ]$.

\subsection{The mass for random combs}

In this subsection we introduce some auxiliary generating functions
and prove the existence of the mass for
random combs.  We assume that we are given a random comb
where the lengths of the teeth are identically and independently
distributed.
For a fixed comb $C$ define a modified two-point function $G_C^0(x;n)$ by
restricting the sum in  \rf{131} to walks that stop the first time they
hit the vertex $n$.  Then we have the factorization            
\beq{a111}            
G_C(x;n)=G_C^0(x;n)Q_C(x;n),            
\eeq
where $Q_C(x;n)$ is the sum over all walks which begin and end at $n$ and
avoid the root $0$.  Equation 
\rf{a111} can be obtained by considering any walk
contributing to the two-point function $G_C(x;n)$ and cutting it at $n$ the
first time it hits $n$.  The first part then contributes to $G_C^0(x;n)$
while the second part contributes to $Q_C(x;n)$.
Let $T_n$ be the tooth of $C$ at $n$ (which may be
empty).  Let $P_C^{(-)}(x;n)$ be the generating function for first return of
random walks that begin at $n$, have a first step to $n-1$, and avoid the the
root.  Similarly let $P_C^{(+)}(x;n)$ be the generating function for first
return of
random walks that begin at $n$ and have a first step to $n+1$.
Then we have            
\beq{a112}            
Q_C(x;n)={\sigma (n) \over 3-P_C^{(-)}(x;n)-P_C^{(+)}(x;n)-P_{T_n}(x)}.            
\eeq
Using $P_C^{(+)}(x;n)\leq 1-\sqrt{x}$, $P_C^{(-)}(x;n)\leq 1$ and $P_{T_n}(x)\leq 1$
we obtain            
\beq{a113}            
Q_C(x;n)\leq 3x^{-\oh}            
\eeq            
and hence,            
\beq{a114}         
G_C^0(x;n)\leq G_C(x;n)\leq 3x^{-\oh}G_C^0(x;n).            
\eeq
Let us define $G_C^0(x;n,n')$ as the sum, analogous to \rf{131}, 
over all walks from $n$ to $n'$
which avoid both $n$ and $n'$ at all intermediate times.  Then
$G_C^0(x;0,n)=G_C^0(x;n)$.
Consider now a walk $\omega$
contributing to the two-point function $G_C^0(x;n_1+n_2)$.
Cut $\omega$ the first time it hits $n_1$.  Cut it again the last time it
leaves $n_1$.  Then we obtain 3 walks, the first of which contributes to
$G_C^0(x;n_1)$, the second one starts and ends at $n_1$, avoiding both $0$
and $n_1+n_2$ and the last one contributes to $G_C^0(x;n_1,n_1+n_2)$.  We
therefore obtain a factorization            
\beq{a115}            
G_C^0(x;n_1+n_2)=G_C^0(x;n_1)R_C(x;n_1)G_C^0(x;n_1,n_1+n_2),            
\eeq
where $R_C(x;n)\leq Q_C(x;n)$.  Hence, by \rf{a113},            
\beq{a116}            
G_C^0(x;n_1+n_2)\leq 3x^{-\oh} G_C^0(x;n_1)G_C^0(x;n_1,n_1+n_2).            
\eeq
Since the teeth are independently distributed we see that the functions
$G_C^0(x;n_1)$ and
$G_C^0(x;n_1,n_1+n_2)$ are also independently distributed.  
We denote the averaged modified
two-point functions by $\bar{G}^0(x;n,n')$.  Then            
\beq{a117}            
\bar{G}^0(x;n_1,n_1+n_2)=\bar{G}^0(x;n_2),            
\eeq            
so            
\beq{a118}      
\bar{G}^0(x;n_1+n_2)\leq 3x^{-\oh}\bar{G}^0(x;n_1)\bar{G}^0(x;n_2),            
\eeq
and the function $\log(3x^{-\oh}\bar{G}^0(x;n))$ is 
subadditive in $n$.  It follows by  
standard arguments that            
\beq{a119}            
-\lim_{n\to\infty}{\log \bar{G}^0(x;n)\over n}=-\inf_n {\log            
\bar{G}^0(x;n)\over n}.            
\eeq
In view of \rf{a114} we conclude that the mass associated with the averaged
two-point function $\bar{G}(x;n)$ exists and is given by  \rf{a119}. 
               
\subsection{Some bounds}

In this subsection we establish bounds which, for the purpose of
calculating $\alpha$ and $d_s$, allow us to ignore walks that
wander too far along the spine.  We denote by $E(x)$ a generic nonnegative 
function of $x>0$ with
the property that there are positive constants $c_1$, $c_2$ and
$\varepsilon$ such that
\beq{a120}
E(x)\leq c_1e^{-c_2x^{-\varepsilon}}.
\eeq
In the following we will let $c, c', c_1$, $c_2$ etc.\ denote
positive constants whose value may change from line to line.  

Let $C$ be a comb and 
define $P_C^{(n)}(x)$ as the contribution to $P_C(x)$ coming from
walks whose maximal distance from the root along the spine is $n$. 
Then
\beq{a123}
P_C(x)=\sum_{n=1}^\infty P_C^{(n)}(x)
\eeq
and we claim that
\beq{a124x}
\sum_{n=N(x)}^\infty P_C^{(n)}(x) =E(x)
\eeq
if
\beq{a121}
G_C(x;N(x))=E(x).
\eeq
This follows from
\bea
\sum_{n=N(x)}^\infty P_C^{(n)}(x) & = & \sigma (N(x))^{-1} G_C^0(x;N(x))^2
Q_C(x;N(x))\nonumber\\
         & = &  \sigma (N(x))^{-1} G_C(x;N(x))^2Q_C(x;N(x))^{-1},
\eea
cf.\ \rf{a111}, \rf{a113} and \rf{a114}.
We conclude, in particular, that if $N(x)\geq x^{-\oh -\varepsilon}$ for some
$\varepsilon >0$, then \rf{a124x} holds for any $C$.

Finally, if we have a comb $C$ such that \rf{a121} holds for 
$C$ and that $C'$ is
another comb which is identical to $C$ up to the vertex $N(x)$ on the spine,
then, by \rf{a111} and \rf{a114},
\bea
\sum_{n=N(x)}^\infty P_{C'}^{(n)}(x) & \leq & 
(G^0_{C'}(x;N(x)))^2 Q_{C'}(x;N(x))\nonumber\\
  & \leq & cx^{-{1\over 2}}(G_C^0(x;N(x)))^2 \nonumber\\
   & = & E(x). \label{129b}
\eea
The estimates 
\rf{a124x} and \rf{129b} will be used repeatedly in this paper.

\section{Combs with 
infinite teeth at random location}
In this section we
consider a random comb for which there is probability $p\in
(0,1)$
that there is an infinite tooth at a vertex on the spine and probability
$q=1-p$ that there is no tooth, i.e., $\mu (0)=p$ and $\mu (\infty )=1-p$ in
the notation of subsection 2.3.
We will show that in this case the spectral
dimension is $3/2$, i.e., the same as for the full comb.   It follows
immediately that any random comb with a nonzero probability of an infinite
tooth at any given vertex has spectral dimension $3/2$.

The strategy of the
proof is to use \rf{a124x} which shows 
that for a given value of $x$ it suffices to consider
walks that do not move beyond a location $N(x)$ on the spine.
Then we use the Rearrangement Lemma to dilute the teeth on
the interval from $0$ to $N(x) = [x^{-\oh -\varepsilon}]$ 
so that they are regularly spaced and we can
obtain an upper bound on $\bar{Q}$ which turns out to be of the same form as
the trivial lower bound on $\bar{Q}$ coming from comparison with the full comb.   

Let $L_0$ denote the distance from the
root to the first (non-trivial) 
tooth and let $L_i$, $i\geq 1$, denote the distance from the $i$th
tooth to the $(i+1)$st tooth.  
Since the $L_i$'s are independently distributed
random variables we
see that
\beq{311}
\pi (\{L_i\leq L : i=0,1,\ldots ,k-1\})=(1-q^L)^{k}.
\eeq
If $r$ is a real number we denote its integer part by $[r]$.
Fix $\varepsilon \in (0,1/8)$, 
choose $k=[x^{-\oh -\varepsilon}]$ and $L=[x^{-\varepsilon}]$.  
Let $\cA_\varepsilon$ be the event that $L_i>L$ for some $i\in\{ 0,1,\ldots ,k\}$.  
Then, by \rf{311}, $\pi (\cA_\varepsilon )=E(x)$. 

Now consider a comb $C\notin 
\cA_\varepsilon$.  
The spacings between the first
$k$ teeth of $C$ are all smaller than or equal to $L$.
By removing all teeth in $C$ except the first $k$ ones and shifting these
suitably away from the root we obtain a comb $C'$ whose teeth have constant
spacing $L_i=L$, $i=0,\ldots ,k-1$.
Hence, by the Monotonicity and Rearrangement
Lemmas we have 
\beq{312}
P_C(x)\leq P_{C'}(x)= P_{C'}^{(1)}(x) +P_{C'}^{(2)}(x),
\eeq
where $P_{C'}^{(1)}(x)$ is the contribution to $P_{C'}(x)$ coming from
paths which 
do not pass through the point $[x^{-\oh-\varepsilon}]$ on the spine and
$P_{C'}^{(2)}(x)$ is the remainder.  By
\rf{a124x} we have $P_{C'}^{(2)}(x)=E(x)$ uniformly for $C\in \cC\setminus
\cA_\varepsilon$.
Moreover, we have
\beq{tre}
P_{C'}^{(1)}(x)= P_{*L}^{(1)}(x) \leq P_{*L}(x),
\eeq
where $*_L$ is the comb with infinite teeth of spacing $L$.
We conclude from \rf{312} and \rf{tre} that
\beq{313}
P_C(x)\leq P_{*_L}(x) +E(x).
\eeq
Using the result
\beq{314}
P_{*_L}(x)\leq 1-cx^{{1\over 4} +{\varepsilon \over 2}} 
\eeq
derived in Appendix 2, we obtain
\beq{315}
P_C(x)\leq 1 - cx^{{1\over 4}+{\varepsilon \over 2}}
\eeq
and it follows that
\beq{316}
Q_C(x)\leq cx^{-{1\over 4}-{\varepsilon \over 2}} 
\eeq
uniformly for $C\notin \cA_\varepsilon$.  Hence,
\bea
\bar{Q}(x) & = & \int_{\cA_\varepsilon} Q_C(x)\,d\pi (C)
+ \int_{\cC\setminus \cA_\varepsilon} Q_C(x)\, d\pi (C) \nonumber\\
 & \leq & x^{-\oh}\pi (\cA_\varepsilon )+cx^{-{1\over 4}-{\varepsilon\over 2}}
 \pi (\cC\setminus \cA_\varepsilon ) \nonumber\\
& \leq & cx^{-{1\over 4}-{\varepsilon\over 2}} .
\eea
It follows that $\alpha \leq {1\over 4} + {\varepsilon \over 2}$ 
and $d_s\geq {3\over 2} 
- \varepsilon$ for any $\varepsilon >0$.  
In view of the lower 
bound \rf{127} we obtain 
\beq{127xxx}
\alpha ={1\over 4},~~~~d_s={3\over 2}. 
\eeq

\section{Combs with finite random teeth}
In this section we will calculate the spectral dimension of random combs
with finite but arbitrarily 
long teeth.  An upper bound on $\bar{P}(x)$ will be obtained
by mimicking the argument for 
the upper bound obtained in the previous section using the fact
that if the teeth are sufficiently long they can be replaced by 
infinitely long teeth up to discrepancies of size $E(x)$.  The
lower bound will be obtained by a convexity argument.

Let $\mu$ be a probability measure on the non-negative integers and set 
$\mu (\ell )=\mu_\ell$.   For simplicity of presentation we 
assume to begin with that we have a power
law distribution $\mu_\ell =c_a\, \ell^{-a}$, where  
$a>1$ since the $\mu_\ell$ sum to 1.
Choose $\varepsilon >0$ and define
\beq{411}
p(x)=\sum_{\ell=[x^{-\oh -\varepsilon}]}^\infty \mu_\ell,
\eeq
for $x>0$.
We shall refer to $p(x)$ as the probability that a tooth is long.  Clearly
$p(x)\to 0$ as $x\to 0$.  More precisely,
\beq{412}
p(x)\sim x^{g(\varepsilon)},
\eeq 
where
\beq{413}
g(\varepsilon )={a-1\over 2}+\varepsilon (a-1).
\eeq

Consider now the random comb defined by $\mu$ and let 
$M=[x^{-g(\varepsilon )
-\varepsilon}]$ and $K=[x^{-\oh -\varepsilon}]$.
Denote by $L_0$ the distance from the root to the first long tooth and by
$L_i$, $i\geq 1$, the distance from the $i$th long tooth to the $i+1$st long
tooth.   Let $\cB_\varepsilon$ be the event that at least one of the
distances $L_0, \ldots ,L_{K-1}$ is greater than $M$.
Since the probability that $L_i >M$ is given by
\beq{414}
(1-p(x))^{M} \leq e^{-Mp(x)},
\eeq
we have
\beq{req}
\pi (\cB_\varepsilon )\leq K e^{-Mp(x)} =E(x).
\eeq

Consider now a comb $C\notin \cB_\varepsilon$. By deleting all teeth from
$C$ except the first $K$ long teeth and shifting these suitably away from
the root we obtain a comb $C'$ whose teeth have a constant spacing $L_i=M$,
$i=0,\ldots ,K-1$.  By the Monotonicity and Rearrangement Lemmas we have 
\beq{416}
P_C(x)\leq P_{C'}(x).
\eeq
Replacing the teeth in $C'$ by infinitely long ones, only changes
$P_{C'}(x)$ by $E(x)$.
This follows from the fact that walks that go more than a
distance $[x^{-\oh-\varepsilon}]$ into the teeth 
only contribute $E(x)$ to $P_C(x)$ uniformly in $C$.
By the same argument as the one leading to \rf{313} we obtain
\beq{418}
P_C(x)\leq P_{*_M} +E(x),
\eeq
uniformly for $C\notin \cB_\varepsilon$.

Let us first 
consider the case $a<2$.  Then we can choose $\varepsilon$ so small that 
$g(\varepsilon )+\varepsilon < 1/2$.  It follows that $M\sim [x^{-\beta}]$ 
with 
$0<\beta <1/2$ so by \rf{b211x} in Appendix 2, 
\beq{yyy}
P_C(x) \leq 1- cx^{{1\over 4}+\oh (g(\varepsilon)+\varepsilon)},
\eeq
for $C\notin\cB_\varepsilon$.
We conclude from the above and \rf{req} that
\beq{420}
\bar{P}(x)\rangle \leq 1-cx^{{1\over 4}+\oh (g_0+\varepsilon)} 
\eeq
for $\varepsilon$ and $x$ sufficiently small,
where we have defined $g_0=g(0)$.
Hence, 
\beq{422}
\alpha\leq {1\over 4}+{g_0\over 2}={a\over 4}.
\eeq
Similarly we conclude that
\beq{422xx}
\bar{Q}(x)\leq cx^{-{a\over 4}}.
\eeq
If $a\geq 2$ the upper bound on $\alpha$ obtained above is replaced by 
the trivial upper bound $\alpha \leq \oh$ coming from the comparison with
the comb with no teeth and similarly for the exponent of $\bar{Q}$.

We now turn to the proof of the lower bound.   We first note that
\beq{423}
\bar{Q}(x)\geq {1\over 1-\bar{P}(x)}
\eeq 
by Jensen's inequality.  It will suffice to find a suitable lower
bound on $\bar{P}(x)$.  
Noting that the 
lengths of the teeth are independently and identically distributed
we take the expectation of \rf{115} to get
\beq{425}
\bar{P}(x)=\sum_{\ell =0}^\infty \mu_\ell \left\br 
{1-x\over 3-P_\ell (x) - P_{C}(x)}\right\kt .
\eeq
Applying Jensen's inequality again we obtain
\beq{426}
\bar{P}(x)\geq {1-x\over 3-\bar{P}_T (x) - \bar{P}(x)}, 
\eeq
where
\beq{427}
\bar{P}_T (x) = \sum_{\ell =0}^\infty \mu_\ell P_\ell (x).
\eeq
By rearranging the inequality \rf{426} and using the fact that 
$\bar{P}(x)\leq 1$ it follows that
\beq{428}
\bar{P}(x) \geq 1-\sqrt{1+x-\bar{P}_T (x)}.
\eeq
Using the expression \rf{120} for $P_\ell (x)$ and the definition \rf{135}
we can write
\beq{431}
1-P_\ell (x)=\sqrt{x}\, {1-e^{-m(x)\ell}\over 1+e^{-m(x)\ell}}
\eeq
where we have put $m(x)=2\,m_\infty (x)$ for convenience.  
It follows that 
\beq{435}
1-\bar{P}_T(x) 
\leq \sqrt{x}\, \sum_{\ell =0}^\infty \mu_\ell (1-e^{-\ell m (x)}).
\eeq
Consider first the case $a>2$.  Then, from \rf{435}, we obtain 
the bound
\beq{yyy1}
1-\bar{P}_T (x) \leq \sqrt{x}\, m(x)\sum_{\ell =0}^\infty \ell \mu_\ell\leq cx.
\eeq
Hence, $\bar{P}(x)\geq 1-c\sqrt{x}$ and we conclude that $\alpha =\oh$ and 
$d_s=1$.  

More generally define
\beq{436}
I_\gamma =\sum_{\ell =0}^\infty \mu_\ell\, \ell ^{\gamma}
\eeq
for $\gamma \geq 0$.   Let
\beq{437}
\gamma_0=\sup\{\gamma\geq 0: I_\gamma <\infty \} .
\eeq
For the power law distribution $\mu_\ell =c_a\, \ell^{-a}$    
we see that $\gamma_0= a-1$.    
For $a \leq 2$ it follows by a similar argument as before,
using 
\beq{438}
1-e^{-z}\leq z^p,~~~z\geq 0, ~~~0\leq p\leq 1,
\eeq
that for $\varepsilon >0$,
\beq{439x}
1-\bar{P}_T (x) \leq c\sqrt{x}\, m(x)^{\gamma_0-\varepsilon}.
\eeq
We conclude from \rf{428} that
\beq{440}
\bar{P}(x)\geq 1- cx^{{1\over 4} (1+\gamma_0-\varepsilon)}
\eeq
so $\alpha \geq a/4$.
Combining this with \rf{422} shows that $\alpha =a/4$ and therefore 
the spectral dimension is given by
\beq{442}
d_s={4-a\over 2}
\eeq
for $1<a\leq 2$.

Let us now consider the general case when $\mu_\ell$ is an arbitrary
probability distribution 
and assume first that $\gamma_0$ defined as above is finite.
For simplicity, let us further assume\footnote{This assumption is not true for arbitrary
probability distributions and in that case the subsequent arguments will have to be
modified.  We will leave this technical point aside.} that there exists
a non-increasing function $g(\varepsilon )$ such that 
$p(x)\sim x^{g(\varepsilon )}$.   Then 
\beq{443}
\lim_{\varepsilon\downarrow 0}g(\varepsilon )=g_0=\oh \gamma_0.
\eeq
In order to prove this let $\gamma>0$ and consider the integral
\bea
J & = & \int_0^1 p(x)x^{-\oh\gamma -1}\,dx\nonumber\\
   & = & \int_0^1 \sum_{\ell =[x^{-\oh -\varepsilon}]}^\infty
   \mu_\ell \, x^{-\oh\gamma -1}\,dx\nonumber\\
   & \leq & \sum_{\ell =1}^\infty \mu_\ell \int _{(\ell+1)^{-{2\over
1+2\varepsilon}}} ^1 x^{-\oh\gamma -1}\,dx .\label{444}
\eea
Doing the $x$ integral above we see that
\beq{445}
J\leq -c +{2\over \gamma}\sum_{\ell =1}^\infty 
\mu_\ell (\ell +1)^{{\gamma\over 
1+2\varepsilon}}
\eeq
which is finite if $\gamma < (1+2\varepsilon )\gamma_0$.
Similarly we can show that
\beq{446}
J\geq -c' + {2\over \gamma}\sum_{\ell =1}^\infty \mu_\ell \, \ell^{{\gamma\over
1+2\varepsilon}}
\eeq
and the right hand side in the above inequality diverges if $\gamma
>(1+2\varepsilon )\gamma_0$.  We conclude that 
\beq{447}
g(\varepsilon )=\oh\gamma_0 (1+2\varepsilon )
\eeq
so $g_0=\oh\gamma_0$.  This, together with the previous arguments, 
proves that $d_s=1$ if $1< \gamma _0 <\infty$ and 
$d_s=(3-\gamma_0)/2$ if $0\leq \gamma_0\leq 1$.  It is easy to check that 
$d_s=1$ for probability distributions which have $\gamma_0=\infty$.

\section{Random trees}
In this section we show how the results and methods 
of the previous sections can be used
to bound the spectral dimension of infinite planar 
random trees.  We begin by recalling some
results about such trees from \cite{bergfinnur}.  
We consider rooted trees where the root has order 1.
There is a measure
$\pi$ on these trees obtained as a limit of the uniform measures on
ensembles of trees with a finite number of vertices.   With respect to this
measure there is with probability 1 a unique infinite simple path
in a random tree $\tau$ which can be viewed as a ``spine'' $N_\infty$
with finite trees attached to the spine.
For simplicity of presentation, let us consider  
trees whose vertices have order 1, 2 or 3.  Then there is at most 
a single rooted
tree $\tau_j$ with root of order 1 attached to each vertex $j\neq 0$ 
on the spine of the
infinite random tree.  It is shown in \cite{bergfinnur} that the trees
$\tau_j$ are identically and independently distributed with the 
probability distribution
\beq{911x}
\rho (\{\tau_j\}) =Z^{-1} 3^{-|\tau_j|}
\eeq
where $|\tau_j|$ denotes the number of links in $\tau_j$ and $Z$ is a
normalization factor (which in this particular case happens to be equal to
$1$).

The spectral dimension for the infinite random tree $d_s^{{\rm tree}}$
is now defined in the same way 
as the spectral dimension for random combs by considering the return
probability to the root for a simple random walk on the random tree.  
We will show that 
\beq{912x}
{5\over 4}\leq d_s^{{\rm tree}} \leq {3\over 2}.
\eeq
The lower bound is obtained by showing 
that the generating function for first
return of random walks to the root on the random tree, $P_{{\rm tree}}(x)$,
is bounded from above
by the corresponding generating function
for random combs whose tooth length
have a power law distribution with exponent $3/2$.
The upper bound is obtained by a convexity argument.  The result
\rf{912x} is in agreement with $ d_s^{{\rm tree}}=4/3$ found in
\cite{tjjw} by different methods.

In order to prove \rf{912x}
we first note that 
\beq{913x}
c_1n^{-{3\over 2}}\leq \rho (\{\tau : |\tau |=n\}) \leq c_2n^{-{3\over 2}},
\eeq
see, e.g., \cite{book}.  By Lemma A and \rf{442}, we see that it is sufficient
for the lower bound in \rf{912x}
to show that among all the finite rooted trees $\tau$ with a given number
$\ell$ 
of links, it is $N_\ell$ which has the largest first return generating
function.  In order to prove this consider a tree $\tau$ with $\ell$
links and let $P_\tau (x)$ be its first return generating function.
Let $v$ be a vertex in $\tau$ at a maximal distance from the root.
Let $\omega$ be the unique simple path in $\tau$ from the root to $v$. 
Then $v$ has order one and $\tau$ can be viewed as a comb with a finite spine
$\omega$ and finite trees (possibly empty) attached to each 
vertex of $\omega$.  Let $v_j$, $j=0,1,\ldots ,n$, be the vertices of
$\omega$, $v_n=v$. Let $t_j$ be the tree attached to $v_j$.  Let $t_{k}$ be
the first non-empty tree we encounter as we move from $v$ along $\omega$
towards the root.  Let $\tau^\prime$ be the tree obtained 
by swapping $t_k$ and
$t_{k+1}$ (which is empty by hypothesis).  By the argument of the
Rearrangement Lemma we deduce that $P_\tau (x)\leq P_{\tau^\prime}(x)$.
Taking now a vertex $v'$ in $\tau^\prime$ at a maximal
distance from the root and repeating the above argument 
we construct a sequence of trees $\tau ^{(i)}$ and for some finite value 
$i=i_0$  we obtain $\tau^{(i_0)}=N_\ell$.

\begin{figure}[thb]
  \begin{center}
      \includegraphics[width=4.2cm]{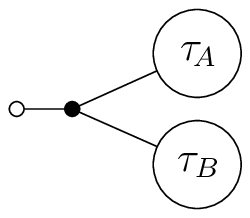}
          \caption{A decomposition of a tree into two subtrees.}
	      \label{figXXX}
	        \end{center}
		\end{figure}

In order to prove the upper bound in \rf{912x} we begin by noting that by
Jensen's inequality
\beq{950}
P_{{\rm tree}}(x)\geq {1-x\over 3-P_{{\rm tree}}(x) -\bar{P}_t(x)}
\eeq
where
\beq{951}
\bar{P}_t(x)={1\over Z}\sum_\tau 3^{-|\tau |}P_\tau (x)
\eeq
and the sum in \rf{951} 
runs over all finite rooted trees with vertices of order 1, 2 or 3.
By the same argument as the one leading to \rf{428} we find that
\beq{952}
P_{{\rm tree}}(x)\geq 1-\sqrt{1+x-\bar{P}_t(x)}.
\eeq
We will now show that 
\beq{980}
\bar{P}_t(x)\geq 1-c\,\sqrt{x}
\eeq
which implies the
desired result.  Given a tree $\tau$ it can always be decomposed into the
root link and two subtrees $\tau_A$ and $\tau_B$ as indicated in Fig.\
\ref{figXXX}.
Note that $\tau_A$ and/or $\tau_B$ may be empty.  This allows us to write
\bea
\bar{P}_t(x) & = & {1\over 3}(1-x)+Z^{-1}
\sum_\tau \left({1\over 3}\right)^{|\tau |+1}
{1-x\over 2-P_\tau (x)}+
\nonumber\\
&& Z^{-2}\sum_{\tau_A, \tau_B} \left({1\over 3}\right)^{|\tau_A |+|\tau_B|+1}
{1-x\over 3-P_{\tau_A} (x) -P_{\tau_B}(x)}.\label{953}
\eea
Using Jensen's inequality again and $Z=1$ we obtain
\beq{954}
\bar{P}_t(x)\geq {1\over 3}(1-x)\left( 1+{1\over 2-\bar{P}_t(x)}
+{1\over 3-2\bar{P}_t(x)}\right).
\eeq
Rearranging \rf{954} leads to
\beq{955}
(6\bar{P}_t(x)-11)(\bar{P}_t(x)-1)^2+
x(2\bar{P}_t^2(x)-10\bar{P}_t(x)+11)\geq 0
\eeq
which implies \rf{980} since $\bar{P}_t(x)\to 1$ as $x\to 0$.

\section{Some non-random combs} 
In this section we use the techniques we have
developed so far to calculate the spectral dimension of some non-random 
combs.  We first remark that it follows easily
from the previous results that any periodic comb has either spectral
dimension $3/2$ (if it has some infinite teeth) or spectral dimension $1$ if
all teeth have finite length.

We will discuss two different types of combs: (i) combs with
infinite teeth where the distance between tooth $n$ and tooth $n+1$ is an
increasing function of $n$ and (ii) combs with finite teeth such that the
length of tooth $n$ is an increasing function of $n$.  
For simplicity of presentation we will choose the
increasing functions to be powers but our methods apply to more general cases.
In both cases we find that the spectral dimension varies continuously with
the power.

\subsection{Combs with increasing tooth spacings.}
Let $a>0$ and define $C$ to be the 
comb all of whose teeth are infinite such that
the distance from the $k$th tooth to the $k+1$st tooth is 
\beq{511}
L_k=[(k+1)^a],
\eeq
where $[r]$ denotes the integer part of $r$ as before. 
The distance from the root to the first tooth is $L_0=1$.
We will show that this comb has spectral dimension
\beq{512}
d_s={3+a\over 2+a}.
\eeq

We first prove an upper bound on $P_C(x)$.  This will be done by an inductive
argument bounding the two point function at a suitable distance along 
the spine:
\beq{bound1}
G_C(x, [x^{-\bar{\eta}-\varepsilon}])=E(x)
\eeq
 for arbtrary $\varepsilon >0$, where 
\beq{niceequation}
\bar{\eta} = {a+1\over 2(a+2)}.
\eeq
For this purpose we need to introduce slightly more general
combs than $C$.  For any $f\geq 1$ let $C^f$ be the
 comb with all 
teeth infinite such that the distance from the
$k$th tooth to the $k+1$st tooth is
\beq{5121}
L_k=[(k+f)^a].
\eeq
Note that $L_0=[f^a]$ in this case and
$C^1=C$.  The inductive hypothesis is as follows. There
exists $\eta_0\in (\bar{\eta},\oh ]$ 
such that for any $\eta>\eta_0$ and any constant $c>0$ there is a
fixed $E$-function, as defined in \rf{a120}, and the inequality
\beq{513}
G_{C^f} (x;[x^{-\eta}])\leq E(x)
\eeq  
holds for all $f$ in the range $1\leq f\leq c x^{-{\eta \over
1+a}}$.
By \rf{a121} the hypothesis is true for $\eta_0=\oh$ since \rf{513} holds
in this case for any value of $f\geq 1$. We shall prove that the
statement then holds also with $\eta_0$ replaced by $\phi (\eta _0)$, where 
\beq{deff}
\phi(\eta )=\frac 14 + \frac{a\eta }{2(a+1)}\,.
\eeq
The strategy of the following argument is to use the induction hypothesis 
\rf{513} to obtain an upper bound on $P_{C^f}(x)$ which in turn will be used
to prove an improved upper bound on $G_{C^f}$ using the representation 
\rf{132}.

Consider one of the combs $C^f$.  The distance from the
root to the $n$th tooth is given by
\beq{514}
D_n=\sum_{k=0}^{n-1}[(k+f)^a]
\eeq
and fulfills
\beq{515}
{1\over a+1}((n+f-1)^{a+1}-(f-1)^{a+1}) \leq D_n\leq {1\over a+1}((n+f)^{a+1}-f^{a+1}).
\eeq
For a fixed $\eta >\eta_0$ choose $n$ as a function of $x>0$, such
that $D_{n+1}\geq [x^{-\eta}]\geq D_n$.  
It follows that 
\beq{5151}
L_n^{-1}\sim x^{{\eta a\over a+1}}
\eeq
since $f^{a+1}\leq c^{a+1} x^{-\eta }$ by assumption.
The contribution to $P_{C^f}(x)$
coming from walks that go beyond $D_{n+1}$ is bounded by $E(x)$
as a consequence of \rf{513}.
By the Rearrangement Lemma and \rf{129b} it follows that
\bea
P_{C^f}(x)& \leq & P_{*_{L_n}}(x)+E(x)\nonumber\\
          & \leq & 1-cx^{\eta '}, 
\eea
where $\eta '=\phi (\eta )$ and we have used \rf{5151} and \rf{b211x} to
obtain the second inequality. 
In order to complete the inductive step we now need to show that
\rf{513} holds with $\eta$ replaced by $\eta '$.
So let $C^f$ be given with $1\leq f\leq 
cx^{-\eta'}$ and let $D=[x^{-\eta '}]$.
By \rf{132} and the Rearrangement Lemma we have
\bea
G_{C^f}(x,D) & = & \sigma (D) (1-x)^{-\oh D}\prod_{k=0}^{D-1}
P_{C^f_k}(x)
\nonumber\\
             & \leq & (1-x)^{-\oh D}(P_{C^{f'}}(x))^D,\label{bigbound}
\eea
where the comb $C^{f'}$ is defined such that the distance from its root to the
first tooth is larger than or equal to the largest
tooth separation up a distance $D$ along the spine in $C^f$.  For this 
purpose it suffices to take  
\beq{5154}
f'=c'x^{-{\eta '\over 1+a}}
\eeq
for a suitable constant $c'$.  The comb $C^{f'}$ is in the class covered by the
induction hypothesis with $\eta$ replaced by $\eta -\varepsilon$ for a
suitable $\varepsilon$ since $\eta '<\eta$.  Hence,
\beq{bound2}
P_{C^{f'}}(x)\leq 1-c_1x^{\phi (\eta -\varepsilon )},
\eeq
and we conclude from \rf{bigbound} that 
\beq{5155}
G_{C^f}(x,[x^{-\eta '}])= E(x)
\eeq
for $1\leq f\leq cx^{-\eta '}$.  
We have thus proven that if the induction hypothesis
 holds for a particular $\eta_0$
it also holds for $\eta_1=\phi (\eta _0)$.
Defining $\eta_r$ inductively by 
\beq{521}
\eta_{r+1}=\phi (\eta_r)
\eeq
we see that $\{\eta_r\}$ is a decreasing sequence in the interval 
$[{a+1\over 2(a+2)},\oh ]$ and
the induction hypothesis holds for all $\eta$ which satisfy
\beq{522}
\eta>\bar{\eta}= \lim_{r\to\infty}\eta_r ={a+1\over 2(a+2)}.
\eeq
We conclude that
\beq{523}
P_C(x)\leq 1-cx^{{1\over 4}+{a\eta\over 2(a+1)}} +E(x)
\eeq
for any $\eta > \bar{\eta}$ and therefore
\beq{524}
P_C(x)\leq 1-cx^{{a+1\over 2(a+2)}+\varepsilon}
\eeq
for any $\varepsilon >0$.  This proves that 
\beq{yyy7}
\alpha \leq {a+1\over 2(a+2)}.
\eeq

In order to prove the corresponding lower bound on $P_C(x)$ 
let $\eta \in (\bar{\eta }, {1+a\over 4})$.  Let
$D_n\sim [x^{-\eta}]$ 
as before and let $C'$ be the comb $C$ with all teeth beyond the
$n$th one removed.  Then from (\ref{129b}) and  (\ref{513}) it follows that
\beq{526}
P_{C'}(x)=P_C(x)+E(x).
\eeq
Define $C(n)$ to be the comb with infinite teeth at the vertices $1,2,\ldots
,n$ but no other teeth.  
\begin{figure}[thb]
  \begin{center}
      \includegraphics[width=7.5cm]{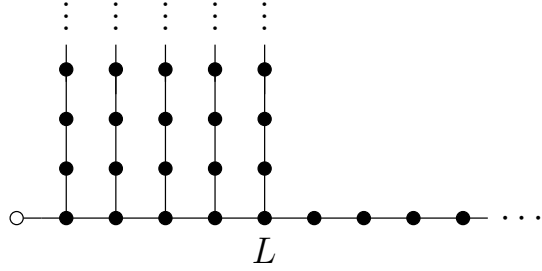}
          \caption{The comb $C(L)$ with $L=5$.}
	      \label{fig3}
	        \end{center}
		\end{figure}
Then, by the Rearrangement Lemma and \rf{526},
\beq{527}
P_{C}(x)\geq P_{C(n)}(x)+E(x).
\eeq
The asymptotic behaviour of $P_{C(n)}(x)$ as $x\to 0$ is computed 
in Appendix 1 and we find
\beq{528}
P_C(x)\geq 1-c\, x^{\oh -{\eta \over a+1}}
\eeq
which implies the desired converse to the inequality \rf{524},
\beq{yyy55}
P_C(x)\geq 1-cx^{\bar{\eta}-\varepsilon}
\eeq
for any $\varepsilon >0$.  We conclude that
\beq{525}
\alpha = {a+1\over 2(a+2)}
\eeq
and the spectral dimension \rf{512} follows.

\subsection{Combs with increasing tooth length}
In this subsection we
consider the comb $C$ for which the length of $T_k$ is given by 
$\ell_k=[k^a], a>0$.  We will prove that the spectral dimension of this comb
is ${3\over 2}$ if $a\geq 2$ but
\beq{529}
d_s={2(1+a)\over 2+a}
\eeq
for $0<a<2$.  In order to prove this we first consider the case $0<a<2$ and 
define the comb $C'$ by
\begin{eqnarray} \ell_k&=&0,\qquad k<k_0\cr\ell_k&=&[k_0^a],\qquad k\ge k_0,
\end{eqnarray}
where $k_0$ will be chosen to depend on $x$ below.
The comb 
$C'$ can be obtained from $C$ by shortening teeth so by the Monotonicity Lemma
\beq{530} P_C (x)\le P_{C'}(x).
\eeq 
We then set $k_0=[x^{-\beta}]$, $0<\beta<\frac{1}{2}$, and choose $\beta$ to 
optimize the bound \rf{530}.  The asymptotic behaviour of the 
function $P_{C'}(x)$ 
is computed in Appendix 1 where this function is denoted $P_{k,\ell}(x)$ with
$k=k_0$ and $\ell =[k_0^a]$.

Taking first the case $\beta a < \frac{1}{2}$ we find
\beq{w111}
P_{C'}(x) = 1-c\, x^{\delta} +O(x^{\delta +\varepsilon}),
\eeq
where $\delta=\max \{\frac{1-\beta a}{2},\beta\}$ and $\varepsilon >0$. 
The value 
of $\delta $ is minimized by choosing $\beta=\beta_{\mathrm{opt}}
\equiv \frac{1}{2+a}$ and we conclude that 
\beq{580}
\alpha \le \frac{1}{2+a},
\eeq
provided that $\beta_{\mathrm{opt}} a< \frac{1}{2}$; that is for $a<2$. 

To obtain a lower bound on $P_C(x)$ we first note that by monotonicity 
$P_{C_k}(x)\le P_C(x)$, as $C_k$ can be obtained from $C$ by lengthening 
teeth, and therefore
\begin{eqnarray} G_C(x;n)&=&(1-x)^{-n/2}\sigma (n)
\prod_{k=0}^{n-1}P_{C_k}(x)\cr
&\le& 3(1-x)^{-n/2} (P_{C}(x))^n\cr
&\le& c_1\,\exp\left(-c_2\,n\, x^{\frac{1}{2+a}}\right).
\end{eqnarray}
Combining this fact with \rf{129b} and the Monotonicity Lemma shows that 
\beq{581}
P_C(x)\ge P_{C''}(x) +E(x),
\eeq
where $C''$ is the comb with teeth of constant length 
$\ell_k=[k_1^a]$ and $k_1=[x^{-{1\over 2+a}-\varepsilon}]$.   
This comb has  exactly the structure of the 
comb $\#$ considered in Appendix 1.  
We find, using \rf{bb139}, 
\beq{582}
P_{C''}(x)\geq 1-c x^{\frac{1}{2+a}-\varepsilon^\prime}
\eeq
for any $\varepsilon^\prime>0$.
It follows that 
$\alpha\ge \frac{1}{2+a}$; combining this with
\rf{580} gives the results
\beq{583}
\alpha=\frac{1}{2+a}, ~~~d_s = \frac{2+2a}{2+a},\quad 0<a<2.
\eeq

We now turn to the case $a\ge 2$.  We use the argument leading to 
\rf{w111} with $\beta ={1\over 4}$.
The teeth of $C'$ are now so long that they are effectively 
infinite and \rf{bb127}, \rf{bb134} and \rf{bb135} yield
\beq{fgh} 
P_C(x)\leq P_{C'}(x)=1 -cx^{{1\over 4}} +O(x^{{1\over 4}+\varepsilon})
\eeq
where $\varepsilon >0$. 
From \rf{127} we know that $P_C(x)\geq P_*(x)$ and it 
follows immediately that 
\beq{585}
\alpha=\frac{1}{4},~~~
d_s=\frac{3}{2}, \quad a\ge 2.
\eeq

\section{Anomalous diffusion} 
In this section we explore the connection between the full heat kernel on
random combs and the functions we have focused on so far, namely the
two-point function and the first return generating function.  The main
result is that anomalous diffusion along the spine is described by the 
decay of the two
point function and the critical exponents $\alpha $ and $\nu$ coincide.
We will focus on the random comb with random toothlength.   The
comb with random spacing between infinitely long teeth can be treated by
similar arguments.

\subsection{The exponents $\alpha$ and $\nu$ are equal}
Our starting point is the representation \rf{132} of the two-point function
on a comb $C$.   We will bound $\nu$ from below with $\alpha$ using a
convexity argument.  The opposite inequality follows from pointwise
estimates as obtained in Sections 3 and 4 in the calculation of the spectral
dimension. 

Using \rf{115} and \rf{132}, and remarking that
\beq{612}
{1\over 1+y}\geq e^{-y}
\eeq
for any $y\geq 0$, we find
\beq{613}
G_C(x;n)\geq \sigma (n) (1-x)^{n/2}\exp\left( 
-\sum_{k=0}^{n-1}(2-P_{T_{k+1}}(x)-P_{C_{k+1}}(x))\right) .
\eeq
Averaging over the comb ensemble and applying Jensen's inequality
we obtain
\beq{614}
\bar{G}(x;n)\geq  (1-x)^{n/2}e^{-n(2-\bar{P}_T(x)-\bar{P}(x))}
\eeq
where $\bar{P}_T(x)$ is the average of the first return generating functions on
the individual teeth.  Clearly $\bar{P}_T(x)\geq \bar{P}(x)$ so
\beq{615}
\bar{G}(x;n)\geq (1-x)^{n/2}e^{-2n(1-\bar{P}(x))}.
\eeq
We conclude immediately that $\nu\geq \alpha$.

In order to prove the converse inequality it is sufficient to show
that 
\beq{616a}
\bar G (x;[x^{-\alpha-\varepsilon'}]) = E(x)
\eeq
for arbitrarily small $\varepsilon'>0$. Indeed, it follows from the
definition of the mass and \rf{a114} that
\beq{bergf1}
\bar G(x;n)\geq \bar G^0(x;n)\geq e^{-m(x)n}
\eeq
for all $n$. Hence \rf{616a} implies
$m(x)x^{-\alpha-\varepsilon'}\to\infty$ as $x\to0$, which shows that
$\nu\leq\alpha+\varepsilon'$. 

To establish \rf{616a}, we split the average over
$\cC$ into a contribution from $\cB_{\varepsilon}$ and a contribution
from $\cC\setminus\cB_{\varepsilon}$ as we did in Section 4 for
$\bar P(x)$. By \rf{137} the former contribution is bounded from above
by 
\beq{berg2}
\frac 32 G_{\infty}(x;n) \pi(\cB_{\varepsilon}) = E(x).
\eeq
For $C\in\cC\setminus\cB_{\varepsilon}$ we recall from Section 4 that
the first return generating function can be estimated by
\beq{616}
P_C(x)\leq P_{*_M}(x)+E(x),
\eeq
where $M=[x^{-g(\varepsilon )-\varepsilon}]$, and $E(x)$ is independent of $C$.
We claim that the bound \rf{616} also holds for
$C_k,\,k = 1,\dots,[x^{-\alpha-\varepsilon'}]$, uniformly in
$\cC\setminus\cB_{\varepsilon}$ for $\varepsilon'$ sufficiently
small. Recalling that combs 
$C\in\cC\setminus\cB_{\varepsilon}$ are charcterized by the
requirement $L_0,\dots,L_{K-1}\leq M$, where $K=[x^{-\frac
12-\varepsilon}]$, and that $\frac 14\leq \alpha\leq \frac 12$, we 
choose $\varepsilon'$ such that $\alpha + \varepsilon' < \frac 12 +
\varepsilon$, and hence $x^{-\alpha - \varepsilon'}/K \to 0$ as $x\to 0$.
In particular, for each $C_k$ under consideration, we have that
$L_0,\dots,L_{[\frac12 K ]}\leq M$, for $x$ sufficiently small, and
inspection of the proof of
\rf{616} shows that the inequality holds for the $C_k$'s as claimed
(with a modified $E$-function). We can thus use \rf{616} together with
the product representation \rf{132} and obtain for the second
contribution to $\bar G (x;[x^{-\alpha-\varepsilon'}])$ the bound
\beq{617}
 (1-x)^{n/2}(P_{*_M}(x)+E(x))^{[x^{-\alpha-\varepsilon'}]}= E(x)\,.
\eeq
We have thus proved \rf{616a} and hence also $\nu\leq \alpha$. 

\subsection{The heat kernel}
Consider a comb $C$ and 
define $K_C(t;n,k)$ as the probability that a random walker who leaves
the root at time $0$ is located at the vertex $k$ in the $n$th tooth of $C$ at
time $t$.  We will denote this vertex as $(n,k)$.  
If the $n$th tooth of $C$ has length smaller than $k$ we define
this probability to be $0$.  We will refer to the function $K_C(t;n,k)$ as
the heat kernel on $C$ since it satisfies the heat (or diffusion) 
equation on $C$:
\beq{heat}
K(t+1;n,k)=\sum_{(n',k')}\sigma (n',k')^{-1}K(t;n',k'),
\eeq
where the sum in \rf{heat} runs over the nearest neighbours of the vertex
$(n,k)$ in $C$.
Next define the function
\beq{620}
K_C(t;n)=\sum_{k=0}^\infty K_C(t;n,k)
\eeq
which is the probability that a walker has travelled a distance $n$ along
the spine at time $t$.  We are interested in the asymptotic behaviour of
$K_C(t;n)$ for large $t$ and $n$.  In order to analyse this function we define
the corresponding Green function by the Laplace transformation
\beq{621}
H_C(x;n)=\sum_{t=0}^\infty (1-x)^{t/2}\,K_C(t;n).
\eeq
Decomposing the walks contributing to the heat kernel we can express
$H_C(x;n)$ in terms of previously defined generating  functions as
\beq{621a} H_C(x;n)={G_C(x;n)\over 1-P_C(x)} D_\ell(x)\eeq
where $\ell$ is the length of the $n$th tooth of $C$ and
\beq{621b}  
D_\ell(x) =  1+{1\over 3}\sum_{k=1}^\ell G_{N_\ell} (x;k).
\eeq
Using \rf{132} and \rf{120} we can write $ D_\ell(x)$ in terms of $P_\ell(x)$ as
\beq{631}
D_\ell (x)= \frac{2}{3} +\frac{1}{3}\left(1+\sqrt{1-x}\right) \left({1-P_\ell (x)\over
{x}}\right). 
\eeq


We begin by establishing a lower bound on $H_C$.   Let 
$C(\infty n)$ denote the comb $C$ with
the $n$th tooth replaced by an infinite tooth.
By the Monotonicity Lemma and \rf{132} it follows that
\beq{aaa}
P_{C(\infty n)}(x)\leq P_C(x),~~~G_{C(\infty n)}(x;n)\leq {3\over 2}G_C(x;n).
\eeq
Hence,
\beq{622}  H_C(x;n)\ge {
(1-x)^{-n/2}D_\ell(x)\over 1-P_{C(\infty n)}(x)}\prod_{k=0}^{n-1}
P_{C_k(\infty n-k)}(x).
\eeq
We note that $D_\ell (x)$ is the only quantity on the right hand side of
the inequality \rf{622} which depends on the length of the $n$th tooth of $C$.
Now consider one of the combs $C_k(\infty n-k)$ and swap teeth $n-k-1$ times 
so that the infinite tooth becomes the first tooth while the tooth $T_j$
of $C_k(\infty n-k)$ becomes tooth number $j+1$ for $j=1,2,\ldots ,n-k-1$.
Denote the resulting comb by $C'_k(\infty n-k)$.
By the Rearrangement Lemma $P_{C_k(\infty n-k)}(x) \geq  
P_{C'_k(\infty n-k)}(x)$.
By \rf{115},
\beq{624}
P_{C'_k(\infty n-k)}(x) =  {1-x\over 3-P_\infty (x) -P_{\tilde{C}_{k+1}}(x)}.
\eeq
where $\tilde{C}=C'(\infty n)$.  We now 
 average over $C$.  Using \rf{612}, combined with \rf{624}, and Jensen's
inequality we obtain,
\beq{626}
\langle H_C(x;n)\rangle\ge {
(1-x)^{n/2}\langle D_\ell(x)\rangle\over 1-\langle P_{C(\infty n)}(x)\rangle }
e^{-n(2-P_\infty (x)-
\bar{P}(x))}.
\eeq
We now take the ensemble average of $D_\ell$.  Let us first assume that the
teeth are finitely long with probability 1 and let $\gamma_0$ be given by
\rf{437}.  Then, by \rf{439x}, 
\beq{632}
1-\langle P_\ell(x)\rangle \leq c\,x^{\oh (1+\gamma_0 -\varepsilon)}.
\eeq
The converse inequality, with $\varepsilon$ replaced by $-\varepsilon$, 
follows from \rf{420} and \rf{440} so
\beq{633}
1-\langle P_\ell(x)\rangle \sim x^{2\alpha }.
\eeq
We conclude from \rf{631} that
\beq{634}
\langle D_\ell (x)\rangle \geq c_1 +c_2x^{2\alpha -1}.
\eeq
By \rf{aaa}, $\langle P_{C(\infty n)}(x) \rangle \leq \bar{P}(x)$.  
On the other hand,
\rf{624} with $k=0$ and Jensen's inequality imply that
\beq{azx}
\langle P_{C(\infty n)}(x)\rangle \geq  {1-x\over 3-P_\infty (x) -
\bar{P}(x)}
\eeq
and we conclude that
\beq{azx2}
1-\bar{P}(x)\sim 1-\langle P_{C(\infty n)}(x)\rangle .
\eeq
Combining \rf{626}, \rf{634} and \rf{azx2} yields
\beq{635}
\langle H_C(x;n)\rangle \geq c_1\,x^{\alpha -1}e^{-c_2nx^\alpha}.
\eeq
In order to establish the corresponding upper bound we use the inequalities
\rf{616} and \rf{617} which imply that
\beq{azx3}
H_C(x;n)\leq {(1-x)^{-n/2} D_\ell (x)\over 1-P_{*M}(x)}\left( (
P_{*M}(x)+E_1(x))^n +E_2(x)(P_\infty (x))^n\right),
\eeq
where the $E$-functions $E_1$ and $E_2$ are uniform in $C$.
It is clear from \rf{631} that
the converse of the inequality \rf{634} holds (with different constants
$c_1$ and $c_2$). Hence, 
\beq{636}
\langle H_C(x;n)\rangle \sim x^{\alpha -1}e^{-c_2nx^\alpha}
+E(x)e^{-n\sqrt{x}}
\eeq
for small $x$.
If infinite teeth appear with nonzero probability in the comb ensemble, then
$\alpha =\oh$ and we can ignore the finite teeth since they do not affect the
critical behaviour.  In this case, \rf{636} holds with $\alpha =\oh$.

The asymptotic relation \rf{636} enables us to compute the mean 
extent down the spine of walks at large time. We have, for $k>0$,
\bea 
\sum_t(1-x)^{t/2}\langle\,\langle n^k
\rangle_{\omega:\vert\omega\vert=t}\rangle_C&=&\sum_n n^k 
\langle H_C(x;n)\rangle\cr
&\sim&{c_1\over x^{1+k\alpha}} +E(x),\quad x\to 0,
\eea
and so, by standard Tauberian theorems (see, e.g., \cite{titchmarsh,feller}),
\beq{636a} \langle\,\langle n^k\rangle_{\omega:\vert\omega\vert=t}\rangle_C\sim c_2 t^{k\alpha},\quad t\to\infty.\eeq
On the other hand, \rf{636} does of course not allow us to compute
$\langle K_C(t;n)\rangle$.  However, if we make the ansatz,
\beq{637}
\langle K_C(t;n)\rangle \approx c\,{n^\delta\over
t^\gamma}e^{-c'n^\beta/t^\epsilon},
\eeq
we find that this is consistent with \rf{636} if and only if
\beq{638}
\beta= {1\over 1-\alpha},~~
\epsilon={\alpha\over 1-\alpha}, ~~\gamma={\alpha\over 2(1-\alpha )},~~
\delta={2\alpha -1\over 2(1-\alpha )}.
\eeq
This is in agreement with the exact results for the half line
\cite{half} and
the full comb \cite{fullcombmath}.   
We believe that \rf{637} captures the
essential behaviour of the averaged heat kernel for 
\beq{639}
{n\over t^\alpha}\gg 1.
\eeq
Of course $K_C(t;n)=0$ for $t<n$ and for $n$ slightly smaller than $t$ it is
clear that $K_C(t;n)$ decays exponentially in $n$.   We also know that
$\langle K_C(t;0)\rangle\sim t^{-d_s/2}$.  We believe that $\langle
K_C(t;n)\rangle $ is a decreasing function of $n$ for a fixed $t$ but a 
proof does not seem to be straightforward.

\subsection{First passage}
Let $q_C(t;n)$ be the probability that a walk which leaves the root at time
$t=0$ on a comb $C$ hits the vertex $n$ on the spine for the first time 
at time $t$.  We define the corresponding generating function as
\beq{640}
U_C(x;n)=\sum_{t=0}^\infty (1-x)^{t/2} q(t;n).
\eeq
The mean first passage time at $n$ is defined as the quantity
\beq{641}
\bar{t}_C(n)=\sum_{t=0}^\infty t\,q_C(t;n)=-2\left.{\pa\over \pa
x}U_C(x;n)\right|_{x=0}.
\eeq
One can calculate the probability $q_{N_\infty}(t;n)$ explicitly by
elementary combinatorics with the result 
\beq{642}
U_{N_\infty}(x;n)={1\over \cosh (m_\infty (x)n)}
\eeq
which implies that the mean first passage time on the half line is
$\bar{t}_\infty (n)=  n^2$.

In this paper we shall not attempt to make a full calculation of the
averaged probability distribution $\langle q_C(t;n)\rangle$ but rather be
content with showing that all the average 
mean first passage times are infinite on
random combs with $\alpha <\oh$. For this purpose it clearly suffices to
show that $\langle \bar{t}_C(2)\rangle$ is infinite.  The physical reason
for this is easily seen to be that if the teeth in a random comb
are sufficiently long to shift the spectral dimension away from $1$ then the
average time spent in a random tooth is infinite.

Let us first assume that we have a random comb where there is a non-vanishing
probability $p$ that we have an infinite tooth at each vertex on the spine.
Then
\beq{643}
\langle \bar{t}_C(2)\rangle \geq \left.-\left({2p\over 9}\right) 
{d\over
dx}P_\infty (x)\right|_{x=0} =\infty 
\eeq
where the lower bound is obtained by taking only into account combs with an
infinite first tooth and restricting the attention to walks that
wander into the first tooth and proceed directly to the vertex $2$ when they
return from the first tooth.  Similarly, if the teeth are finite with
probability distribution $\mu_\ell$,
\beq{644}
\langle \bar{t}_C(2)\rangle \geq \left.- {2\over 9}{d\over dx}\sum_{\ell
=1}^\infty \mu_\ell P_\ell (x)\right|_{x=0}.
\eeq
Using \rf{120} 
the right hand side in \rf{644} is easily seen to be bounded from below by
\beq{645}
c\,\lim_{x\to 0}\sum_{\ell =0}^\infty \mu_\ell \ell e^{-m_\infty (x)\ell}
\eeq
which is infinite 
if $\alpha =(\gamma_0 +1)/4 <1/2$ where $\gamma_0$ is defined by \rf{437}.

\section{Discussion}
Heat kernels on graphs and Riemannian manifolds have been extensively
studied by mathematicians, see, e.g., \cite{grigoryan,coulhon} and
references therein.  Much of this work is aimed at establishing the connection
between pointwise behaviour of the heat kernel and geometrical properties of
the graphs and manifolds.  The most relevant results from our point of view
are inequalities which, in our notation, are written
\beq{711}
{2d_H\over 1+d_H}\leq d_s\leq d_H,
\eeq
valid for graphs where the Hausdorff dimension exists and is finite, see
\cite{grigoryan}  Theorems 2.2 and 2.3.  The
Hausdorff and spectral dimensions calculated in this paper are readily seen
to satisfy these bounds, some saturate the lower bound, others saturate
the upper bound and some are in between, see Table 1.   

It is not understood in detail which properties of a graph cause the
spectral and Hausdorff dimensions to differ.  For the special and rather
simple example of combs it is easy to see that if we have mostly short
teeth, then the spectral and Hausdorff dimensions are both equal to $1$.  As
the teeth grow, both dimensions grow but the Hausdorff dimension grows
faster in general.
It would be interesting
to relate the two dimensions to the distribution of the orders of vertices.
For more general graphs the connectivity clearly plays a role, not 
only the order distribution.

\bigskip

\renewcommand{\arraystretch}{1.3}
$$
\begin{array}{|l |c|c|c|}
\hline {\rm } & d_H & d_s & {2d_H\over 1+d_H}\\
\hline \mbox{\rm Random tooth spacing} & 2 & {3\over 2} & {4\over 3}\\
\hline
\mbox{\rm Random tooth length}~~a\geq 2 & 1 &1 &1\\
\hline 
\mbox{\rm Random tooth length}~~1<a<2  & 3-a & {4-a\over 2} & {6-2a\over 4-a}\\
\hline
\mbox{\rm Growing tooth spacing} & {2+a\over 1+a} & {3+a\over 2+a} &
{4+2a\over 3+2a}\\
\hline
\mbox{\rm Growing tooth length}~a\geq 2 & 2 & {3\over 2} & {4\over 3}\\
\hline
\mbox{\rm Growing tooth length}~1\leq a<2 & 2 & {2(1+a)\over 2+a} & {4\over
  3}\\
\hline
 \mbox{\rm Growing tooth length}~0<a<1 & 1+a & {2(1+a)\over 2+a} & {2+2a\over
   2+a} \\
\hline 
\mbox{\rm Random trees} & 2  & {4\over 3} & {4\over 3}\\
\hline
\end{array}
$$

\bigskip

\centerline{{\bf Table 1.}  The spectral and Hausdorff dimensions discussed
in this paper.}

\bigskip

There is no analytical understanding of the spectral dimension of random
surfaces and higher dimensional random manifolds.  The spectral dimension of
random surfaces is believed to be $2$ \cite{earlynumsim,scaling} 
while the Hausdorff dimension is known to be $4$ \cite{ambjornwatabiki}, see
also \cite{angel1,angel2}.  It has been shown recently 
\cite{angel1,angel2} that the generic 
structure of infinite planar random surfaces (triangulations) is 
analogous to that of the
random infinite trees discussed in Section 5.  If we take such a 
surface $S$ with a marked vertex and look at the boundary of a ball $B$, 
the boundary will have a number of disjoint components and with probability
1 only one of these components bounds an infinite subsurface of $S$.
This means that we can view the infinite planar random surface as a tube
with finite size outgrowths (baby universes) which are in fact 
distributed in a
simple way analogous to \rf{911x}.  The tube and the outgrowths on the
random surface  
correspond to the spine and the teeth of the random comb.  Whether this
picture allows us to obtain a rigorous control over the spectral dimension
of random surfaces remains to be seen.

\bigskip

\noindent
{\bf Acknowledgements} 

This work is supported in
part by MatPhySto funded by the Danish National Research Foundation and 
by TMR grant no. HPRN-CT-1999-00161.  T.~J. and J.~W. are grateful for 
hospitality at the University of Copenhagen where this work was begun.  
T.~J. would like to acknowledge hospitality at U.~C. Santa Barbara and
Rutgers University.

\newpage

\noindent
{\Large\bf Appendix 1}

\medskip

In this appendix we compute the
first return generating functions for a number of simple combs.  The method
is essentially to solve the recursion relation \rf{115} in simple cases.

We begin by evaluating the first return generating function $P_\ell (x)$ for
finite teeth.
We note that $P_1(x)=1-x$ and
\beq{bb111}
P_\ell (x)={1-x\over 2-P_{\ell -1}(x)}.
\eeq
Let us define 
\beq{bb11x}
\Delta_\ell = {P_\ell -P_\infty\over 2-P_\infty}.
\eeq
It follows from \rf{bb111}
that $\Delta_\ell$ satifies the recursion relation
\beq{bb112}
\Delta_\ell=\left({P_\infty\over 2-P_\infty}\right){\Delta_{\ell -1}\over
1-\Delta_{\ell -1}}.
\eeq
Writing
\beq{bb113}
A={2-P_\infty\over P_\infty} ~~{\rm and}~~ X_\ell=\Delta_\ell^{-1}
\eeq
we see that  \rf{bb112}
can be written
\beq{bb114}
X_\ell=A(X_{\ell -1} -1)
\eeq
which has the solution
\beq{bb115}
X_{\ell+1}=A^\ell X_1-A{1-A^{\ell }\over 1-A}.
\eeq
Inserting the values of $A$ and $X_1$ and doing some 
algebra leads to the desired
result \rf{120}.

Next, let us consider the comb 
$C(L)$ which has infinite teeth at $1,2,\ldots ,L$ but no other teeth.
Then, by \rf{115},
\beq{bb116}
P_{C(L)} (x)={1-x\over 3-P_\infty -P_{C(L-1)}(x)}.
\eeq
Defining 
\beq{bb117}
E_L={P_{C(L)}-P_*\over 3-P_\infty-P_*}
\eeq
we find from \rf{bb116} that $E_L$ satisfies the recursion relation
\beq{bb118}
E_L={P_*\over 3-P_\infty -P_*}\, {E_{L-1}\over 1-E_{L-1}}
\eeq
which is of the same form as \rf{bb112}.   Hence, by the same reasoning as
that leading to \rf{bb115} we find that
\beq{bb119}
E_L={1\over B^LE_0^{-1}-B{{\displaystyle {1-B^L\over 1-B}}}}
\eeq
where 
\beq{bb120}
B={3-P_\infty-P_*\over P_*}.
\eeq
Hence,
\beq{bb121}
P_{C(L)}-P_*={P_\infty-P_*\over 1+F(B^L-1)},
\eeq
with
\beq{bb122}
F=1+ {P_\infty -P_*\over P_* (B-1)}.
\eeq
Noting that $B=1+2x^{1/4}+O(\sqrt{x})$ as $x\to 0$ we see that
\beq{bb123}
F={3\over 2}+O(x^{1/4}).
\eeq
The comb $C(L)$ clearly has spectral dimension 1 since it only has a finite
number of teeth.  However, for our application in Section 6, we are interested
in the behaviour of $P_{C(L)}(x)$ as $x\to 0$ with $L$ behaving like a negative 
power of $x$. 

Let us now assume that $L = [x^{-\beta}]$ with $0<\beta<{1\over 4}$.
It follows by expanding out the denominator in \rf{bb121} that
\beq{bb124}
P_{C(L)}(x)=1-3x^{\oh -\beta }+o(x^{{1\over 2}-\beta }).
\eeq
We conclude that
\beq{bb125}
1-P_{C(L)}(x)\sim x^{\oh -\beta}
\eeq
which was needed for the inequality \rf{528}.

The final comb we consider in this section has no teeth at vertices
$0,1,\ldots k-1$ but teeth of length $\ell $ at $k,k+1, \ldots $.  
Denote this comb by $C(k,\ell)$.
\begin{figure}[thb]
  \begin{center}
      \includegraphics[width=7.5cm]{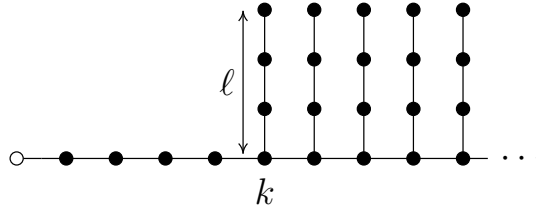}
          \caption{The comb $C(k,\ell )$ with $k=5$ and $\ell =4$.}
	      \label{fig4}
	        \end{center}
		\end{figure}
We will
be interested in the limit of the first return generating function on 
$C(k,\ell )$, denoted $P_{k,\ell}$, 
when $k$ and $\ell$ tend to infinity as $x\to 0$.    
The logic of the calculation is very much the same as above.

Let us denote $P_{1,\ell}$ by $P_\#$ and note that this function satisfies the
recursion relation
\beq{bb126}
P_\#(x)={1-x\over 3-P_\ell (x)-P_\#(x)}
\eeq
which is easily solved with the result
\beq{bb127}
P_\#(x)={3-P_\ell(x)\over 2}-\sqrt{1-P_\ell (x)+{1\over 4}(1-P_\ell (x))^2
+x}\; .
\eeq
With $A$ defined as in \rf{bb113} one finds
\beq{bb128}
{2-P_\infty\over P_{k,\ell}-P_\infty}= A^{k-1}{2-P_\infty\over P_\#-P_\infty }
-A{1-A^{k-1}\over 1-A},
\eeq
cf.\ \rf{bb115}.  This can be written
\beq{bb129}
P_{k,\ell}-P_\infty = {2(P_\#-P_\infty )(1-P_\infty )\over
2(1-P_\infty )+(2-P_\infty -P_\# )(A^{k-1}-1)}.
\eeq
We now wish to find the asymptotic behaviour of $P_{k,l}(x)$ as $x\to 0$
with $k=[x^{-\beta}]$ and $\ell =[k^a]$.  Assume that $0<\beta <\oh$ and
recall that
\beq{bb130}
A={1+\sqrt{x}\over 1-\sqrt{x}}.
\eeq
It follows that
\beq{bb131}
A^k-1=2\, x^{\oh -\beta} +o(x^{\oh-\beta}).
\eeq
Note that \rf{120} can be written
\beq{bb132}
1-P_\ell (x)=\sqrt{x}\,{A^\ell -1\over A^\ell +1},
\eeq
so if $\beta a<\oh$ and we use 
\rf{bb131} with $\beta$ replaced by $\beta a$ we find
\beq{bb133}
1-P_\ell (x)=x^{1-\beta a}+o(x^{1-\beta a}).
\eeq
In the case $\beta a >\oh$ it is not hard to see that
\beq{bb134}
1-P_\ell (x)=\sqrt{x}+E(x).
\eeq
In the limiting case $\beta a =\oh$ an easy calculation yields
\beq{bb135}
1-P_\ell (x)={e^2-1\over e^2+1}\,\sqrt{x}+o(\sqrt{x}).
\eeq
The next step is to find how $P_\# (x)$ behaves as $x\to 0$ and then we can
infer the behaviour of $P_{k,\ell}(x)$ from \rf{bb129}.  It is convenient to
split the argument into two cases.

\medskip
\noindent
{\bf Case 1: {\boldmath $\beta a\geq \oh$}}

\medskip

It is clear from \rf{bb127} that $P_\#(x) =1-c\,x^{{1\over 4}} +o(x^{{1\over
4}})$. 
Using this we find that
\beq{bb136}
P_{k,\ell}(x)-P_\infty (x)=-{cx^{{1\over 4}+\beta } +o(x^{{1\over 4} +\beta})\over 
x^\beta  +cx^{{1\over 4}} +o(x^{{1\over 4}})}.
\eeq
If $\beta < {1\over 4}$ then we find
\beq{bb137}
P_{k,\ell}(x)=P_\infty(x)-cx^{{1\over 4}}+o(x^{{1\over 4}})
\eeq
and if $ \beta >{1\over 4}$ then
\beq{bb138}
P_{k,\ell}(x)=P_\infty(x)-x^{\beta }+o(x^{\beta }).
\eeq
In the crossover case $\beta ={1\over 4}$ 
we find \rf{bb137} with the constant $c$ replaced by 
$c(1+c)^{-1}$.
We remark that this calculation is insensitive to the value of $a$.

\medskip
\noindent
{\bf Case 2: {\boldmath $\beta a<\oh $}}

\medskip
Using \rf{bb127} and \rf{bb133} we obtain
\beq{bb139}
P_\#(x)=1-x^{{1-\beta a\over 2}} +o(x^{{1-\beta a\over 2}}).
\eeq
It follows that
\beq{bb140}
P_{k,\ell}(x)-P_\infty (x)=-{x^{{2-\beta a\over 2}}+o(x^{{2-\beta a\over 2}})
\over \sqrt{x}+x^{1-\beta -\oh \beta a}(1+o(1))}.
\eeq
There are again two cases to consider.  If 
\beq{bbbb140x}
\oh \leq 1-\beta -\oh\beta a
\eeq
then 
\beq{bb141}
P_{k,\ell}(x)=P_\infty(x)-cx^{\oh (1-\beta a)} + o(x^{\oh (1-\beta a)})
\eeq
and $c=\oh$ if equality holds in \rf{bbbb140x} but $c=1$
otherwise.  If $\oh > 1-\beta -\oh\beta a$ then
\beq{bb141xc}
P_{k,\ell}(x)=P_\infty(x)-x^{\beta } + o(x^{\beta}),
\eeq
i.e.,
\beq{bb141cv}
P_{k,\ell}(x)=P_\infty(x)-cx^{\delta } + o(x^{\delta })
\eeq
where $\delta =\max\{\oh(1-\beta a),\beta\}$.  

\bigskip
\noindent
{\Large\bf Appendix 2}

\medskip

In this appendix we show that for $L\sqrt{x}$ small
\beq{b211}
P_{*_L}(x)=1-{x^{{1\over 4}}\over \sqrt{L}}+ O(\sqrt{x}),
\eeq
and hence, for $L=[x^{-\beta}]$ with $\beta <{1\over 2}$, we obtain
\beq{b211x}
P_{*_L}(x)=1-x^{{1\over 4}+\oh\beta}+o(x^{{1\over 4}+\oh \beta})
\eeq
which is the result needed in Sections 3 and 6.

Let $R_L$ denote the generating function for first return to the root for
walks that do not move beyond the $(L-1)$st vertex on the spine, i.e., they
do not reach the vertex where the first tooth appears.  Let $\Gamma_L$
denote the two-point function defined by the sum over all walks from the
root to $L$ which do not return to the root and stop the first time they
meet $L$, i.e., $\Gamma _L(x)=G_C^0 (x;L)$ where $C=N_\infty$.  
Then 
by decomposing the walks that contribute to $P_{*_L}$
we obtain the equation
\beq{b212}
P_{*_L}=R_L+{\Gamma_L^2 \over 3-P_\infty-P_{*_L}-R_L}.
\eeq
It is straightforward to solve this to obtain
\beq{b213}
P_{*_L}={3-P_\infty\over 2}\pm {1\over
2}\sqrt{(3-P_\infty)^2-4(\Gamma_L^2+R_L(3-P_\infty-R_L))}.
\eeq
Since $P_{*_L}(x)\leq 1$ we must choose the $-$ sign in \rf{b213}.
The calculation
of $R_L$ is similar to that of $P_\ell$ in Appendix 1 and gives
\beq{b214}
R_L(x)=(1-x){(1+\sqrt{x})^{L-1}-(1-\sqrt{x})^{L-1}\over
(1+\sqrt{x})^{L}-(1-\sqrt{x})^{L}}
\eeq
so in particular 
\beq{b215}
R_L(x)={L-1\over L}+O(xL).
\eeq
Using an argument similar to the one leading to \rf{132} we see that
\beq{b216}
\Gamma_L(x)=(1-x)^{-\oh L+1}R_{L}(x)R_{L-1}(x)\ldots R_2(x)
\eeq
for $L\geq 2$.  If $L=1$  then the product of the $R$ factors in 
\rf{b216} is equal to 1 by definition.
Rearranging we find
\beq{b217}
\Gamma_L(x)=(1-x)^{L/2}{2\sqrt{x}\over (1+\sqrt{x})^L-(1-\sqrt{x})^L}
\eeq
so
\beq{217x}
\Gamma_L={1\over L}+O(Lx).
\eeq
Noting that the expression under the square root in  \rf{b213} can be written
\beq{b218}
(3-P_\infty -2R_L+2\Gamma_L)(3-P_\infty -2R_L-2\Gamma_L),
\eeq
we obtain \rf{b211} by inserting \rf{119}, \rf{b214} and \rf{217x} into
\rf{b213}. 

\end{document}